\documentclass[aps,pra,twocolumn,superscriptaddress,notitlepage,nofootinbib,longbibliography]{revtex4-1}
\usepackage{mathrsfs}
\usepackage{epsfig}
\usepackage{graphicx}
\usepackage{amsfonts}
\usepackage[figuresright]{rotating}
\usepackage{amssymb}
\usepackage{amsmath}
\usepackage{dcolumn}
\usepackage{bm}
\usepackage{color}
\usepackage{braket}
\usepackage{units}
\usepackage[colorlinks=true, allcolors=blue]{hyperref}

\newcommand{\WQC} {Wilczek Quantum Center, School of Physics and Astronomy, Shanghai Jiao Tong University, Shanghai 200240, China}
\newcommand{\SRCQC}{Shanghai Research Center for Quantum Sciences, Shanghai 201315, China}
\newcommand{\USTCCAS}{CAS Center for Excellence and Synergetic Innovation Center in Quantum Information and Quantum Physics, University of Science and Technology of China, Hefei, 230026, China}
\newcommand{\USTC}{National Laboratory for Physical Sciences at Microscale, University of Science and Technology of China, Shanghai Branch, Shanghai 201315, China}
\newcommand{\UC}{Institute for Quantum Science and Technology, University of Calgary, Alberta, Canada T2N 1N4}
\begin{document}
\title{Sub-Planck structures: Analogies between the Heisenberg-Weyl and SU(2) groups}

\author{Naeem Akhtar}
\email{naeemakhatr@mail.ustc.edu.cn}
\affiliation{\USTC}
\affiliation{\WQC}

\author{Barry C. Sanders}
\affiliation{\USTC}
\affiliation{\USTCCAS}
\affiliation{\UC}

\author{Carlos Navarrete-Benlloch}
\email{corresponding author: derekkorg@gmail.com}
\affiliation{\WQC}
\affiliation{\SRCQC}
\date{\today}

\begin{abstract}
Coherent-state superpositions are of great importance for many quantum subjects, ranging from foundational to technological, e.g., from tests of collapse models to quantum metrology. Here we explore various aspects of these states, related to the connection between sub-Planck structures present in their Wigner function and their sensitivity to displacements (ultimately determining their metrological potential). We review this for the usual Heisenberg-Weyl algebra associated to a harmonic oscillator, and extend it to find analogous results for the $\mathfrak{su}(2)$ algebra, typically associated with angular momentum. In particular, in the Heisenberg-Weyl case, we identify phase-space structures with support smaller than the Planck action in both Schr\"{o}dinger-cat-state mixtures and superpositions, the latter known as compass states. However, as compared to coherent states, compass states are shown to have $\sqrt{N}$-enhanced sensitivity against displacements in all phase-space directions ($N$ is the average number of quanta), whereas cat states and cat mixtures show such enhanced sensitivity only for displacements in specific directions. We then show that these same properties apply for analogous SU(2) states provided (i) coherent states are restricted to the equator of the sphere that plays the role of phase space for this group, (ii) we associate the role of the Planck action to the size of SU(2) coherent states in such a sphere, and (iii) we associate the role of $N$ with the total angular momentum.
\end{abstract}

\maketitle

\section{Introduction}\label{sec:introduction}

The description and generation of quantum states presenting non-classical features has been a recurrent topic since the early days of quantum mechanics, attracting more and more attention as our ability to control quantum coherently optical, atomic, and solid-state systems has developed. Nonclassical properties come in many flavors such as squeezing~\cite{MankoBook03,Schnabel2017}, entanglement~\cite{K2009}, Wigner negativity~\cite{Nonclassical1995,Kenfack2004}, $P$-function divergence~\cite{RICH2002,Cerf1}, and phase-space interference~\cite{BUZEK19951,Gerry05book,sch1}, to name a few. The latter is the subject of our present work, specifically through states built as coherent-state superpositions.
Such states are relevant to many areas, ranging from the foundations of quantum physics to the design and implementation of modern quantum technologies. For example, they can be used to test collapse models aimed at explaining the quantum-to-classical transition~\cite{Collapse1,Collapse2,Collapse3,Collapse4}, but also as resources for quantum sensors with unprecedented resolution~\cite{Zurek2001,Toscano06,Eff4}.

Coherent-state superpositions are especially well studied for the harmonic oscillator~\cite{Gerry05book}, whose position and momentum operators form the so-called Heisenberg-Weyl (HW) algebra~$\mathfrak{hw}(1)$~\cite{weyl1950theory} for a single degree of freedom, and act on an infinite-dimensional Hilbert space. Quantum mechanical states can be visualized on phase space, which is a symplectic manifold, via the Wigner quasi-probability density function~\cite{sch1,CarlosNB15}. As the product of the uncertainties of position and momentum is bounded below by the Planck action $\hbar$, for quite some time phase-space features below this scale were believed not to play a physical role~\cite{Berry79,Korsch81}. Indeed, this is the case for Gaussian states \cite{Cerf2,Cerf3} (coherent, squeezed, thermal, etc.) and even for Schr\"{o}dinger-cat states (superposition of two distinct coherent states~\cite{YS86} or other Gaussian states such as squeezed states~\cite{San89A}),
which show fast oscillations in one direction of phase space, but an unlimited Gaussian profile in the orthogonal direction.

This notion was challenged by Zurek~\cite{Zurek2001}, who showed that the Wigner function of chaotic systems typically develops spotty structures with features below the Planck scale, arguing that these are crucial in determining the sensitivity of the system to decoherence~\cite{ZurekRevMod03} and to phase-space displacements~\cite{Toscano06,Eff4} (ultimately determining its potential for quantum metrology). As prototypical states that show such sub-Planck features,
he built the so-called \textit{compass} states
(one coherent state, so to speak, 
at the north, south, east and west corners), a superposition of four distant coherent states, which can also be understood as superpositions of two cat states. By now, there are multiple theoretical proposals for the controlled generation of these states~\cite{Prop1,Prop2,Prop3,Prop4,Prop5,Prop6}, apart from actual experimental implementations~\cite{Exp1,Exp2,Exp3,Exp4,Exp5}, and their properties and effects in different contexts have been well explored~\cite{Eff1,Eff2,Eff3,Eff5,Eff6,Eff7,Eff8,Eff9,Eff10,Eff11,Eff12,Eff13,Eff14,Eff15}.

The concept of coherent states can be generalized to arbitrary groups~\cite{perelomov1986,Perelomov72}.
Of special interest for physics is the Lie SU(2) group~\cite{Radcliffe1971,Perelomov72,GILMORE1972391,Arecchi72,perelomov1986}, associated with generalized rotations and generated by the angular momentum operators, which satisfy the so-called $\mathfrak{su}(2)$ algebra.
This group includes spin-$\nicefrac12$ whereas the Lie group SO(3) is for integer-labeled rotations.
The space of classical configurations corresponds in this case to the surface of a unit sphere~\cite{Sanders89}.
Superposing two coherent states pinned at antipodal points on the sphere,
one generalizes the cat states first introduced for the~$\mathfrak{hw}(1)$ algebra~\cite{Sanders89,Huang15,Huang18,Maleki,davis2020}.
The cat with states pinned at the sphere's poles is the most studied one, since it corresponds to the popular entangled states known as Greenberger-Horne-Zeilinger (GHZ)~\cite{GHZoriginal} or NOON~\cite{MetrologyDowling08} states depending on the physical implementation, respectively, two-level atomic ensembles or two bosonic modes~\cite{Sanders2014,davis2020}. Even though the Hilbert space on which the $\mathfrak{su}(2)$ algebra acts has finite dimension ($2j+1$ when considering a specific irreducible representation with fixed total angular momentum $j$), one can define a Wigner function that allows visualizing quantum states as quasi-probability distributions defined over the unit sphere~\cite{Varilly89,Heiss00,Klimov17,Glaser20}. The Wigner function of a coherent state appears as a single lobe around the location where it is pinned, with a slight negativity and an effective support that decrease as $j$ increases~\cite{davis2020}. Cat states show an additional interference pattern along the great circle halfway between the locations of their underlaying coherent states~\cite{davis2020} (e.g., the equator for coherent states at the poles). The use of these states for quantum metrology has been analyzed in great depth~\cite{Maccone2004,MetrologyDowling08,Maccone2011,MetrologyRMP18}.

The goal of our present work is two-fold. First, for the~$\mathfrak{hw}(1)$ algebra, we point out that the same sub-Planck structures present in compass states, are present in cat mixtures as well.
Hence, it is argued that sub-Planck structures alone cannot be responsible for the remarkable sensibility of compass states against displacements in \textit{arbitrary} directions of phase space, which is enhanced by~$\sqrt{N}$ with respect to that of coherent states ($N$ is the average number of quanta in the state).
Indeed, it is shown that cat states and mixtures show this enhanced sensitivity only for displacements along \textit{specific} phase-space directions. These results have been discussed in previous literature~\cite{Zurek2001,Toscano06,Eff12} (but without reaching some of the insights that we offer), so this part of the work is to be taken mainly as a review.

The second and main goal of our work is the generalization of these results to the SU(2) group. In particular, we show that the concept of sub-Planck structures can be extended to this case by associating the effective support of coherent states in the sphere, which scales as $\nicefrac1{j}$, with the SU(2) counterpart of the Planck action. Hence,
it is shown that considering coherent-state superpositions along the equator, we can build cat states, compass states, and cat-state mixtures that show similar Wigner interference features as their HW counterparts when represented in the stereographic plane. Finally, we prove that these states have exactly the same enhanced sensitivity to displacements found for the~$\mathfrak{hw}(1)$ algebra, with $j$ playing the role of~$N$.

The organization of this paper follows the structure presented in the preceding paragraph. In~\S\ref{sec:HW} we review the basic concepts for  the~$\mathfrak{hw}(1)$ algebra, which are then generalized in~\S\ref{sec:SU(2)} to the $\mathfrak{su}(2)$ algebra. In~\S\ref{sec:conclusions} we summarize and present our conclusions.

\section{Sub-Planck structures in phase space}
\label{sec:HW}
Let us start by introducing in this section the main concepts that  accompany us throughout this article, including phase-space representations of quantum states, sub-Planck structures, and the metrological potential of coherent-state superpositions. We do this by means of the common example of the harmonic oscillator or~$\mathfrak{hw}(1)$ algebra,
defined through an annihilation operator $\hat{a}$ that satisfies the canonical commutation relation $[\hat{a},\hat{a}^\dagger]=1$. We  work with dimensionless versions of position $\hat{x}:=\hat{a}^\dagger+\hat{a}$ and momentum $\hat{p}:=\text{i}(\hat{a}^\dagger-\hat{a})$,
so-called quadratures, which satisfy the uncertainty relation $\Delta x\Delta p\geq1$, where
\begin{equation}
\label{eq:DeltaA2}
\Delta A^2:=\braket{\hat{A}^2}-\braket{\hat{A}}^2,
\end{equation}
defines the uncertainty of any operator~$\hat{A}$~\cite{Robertson1929, wheeler2014}.

\begin{figure*}[t]
\includegraphics[width=\textwidth]{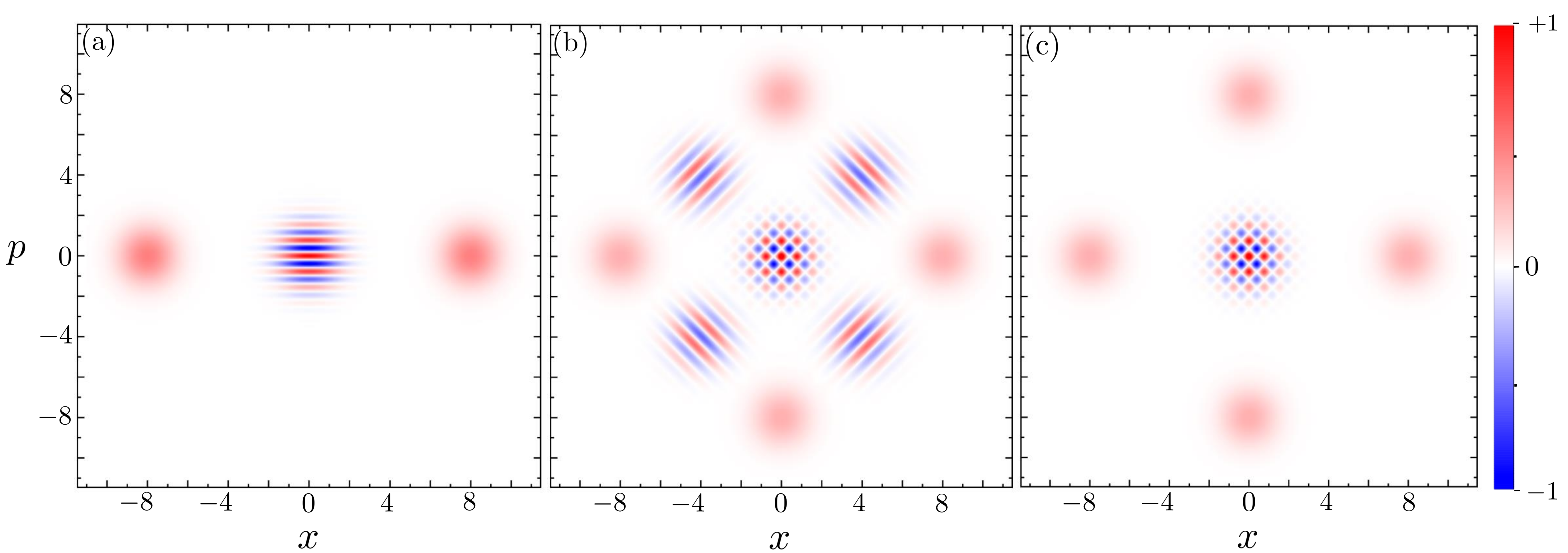}
\caption{Wigner functions of the (Heisenberg-Weyl) coherent-state superpositions considered in this work:
(a)~horizontal cat state,
(b)~compass state, and (c)~cat-state mixture. Red and blue regions correspond, respectively, to positive and negative values of the function, which we normalize to its maximum $W_{\hat{\rho}}(0)$. In all cases we take $x_0=8$.}
\label{fig:Wigner_HW}
\end{figure*}
\subsection{Sub-Planck structures: \\compass state versus cat mixture}\label{subsec:sub-Planck_structures_HW}

Coherent states were introduced by Schr\"{o}dinger as non-spreading wave packets of the quantum harmonic oscillator~\cite{Sch35}.
They are defined~\cite{Glauber63,Gerry05book,CarlosNB15} as the eigenstates of the annihilation operator $\hat a \ket{\alpha}=\alpha\ket{\alpha}$, with $\alpha\in\mathbb{C}$, and are obtained from the vacuum state by acting on the displacement operator $\hat{D}(\alpha)$ as
\begin{align}
	\ket{\alpha}=\hat{D}(\alpha)\ket{0},
	\end{align}
where
\begin{equation}
	\hat{D}(\alpha):=\exp(\alpha\hat{a}^\dagger-\alpha^*\hat{a}).
\end{equation}

In this work we are interested in states built as superpositions of coherent states
\begin{equation}\label{eq:coherent_superposition}
\ket\psi
=\sum_n\psi_n \ket{\alpha_n},~\psi_n \in \mathbb{C}.
\end{equation}
The superposition of two coherent states with the same amplitude and opposite phase leads to the so-called cat state~\cite{DODONOV1974597,Gerry05book}, while the superposition of four coherent states along the same phase-space line is known as the quantum tetrachotomous state~\cite{PhysRevA.99.063813}. On the other hand, superposing four coherent states with the same amplitude and maximally-spread phases we obtain the so-called compass states~\cite{Zurek2001}. We  provide specific examples shortly.

The Wigner function provides a useful way to visualize all these states in phase space~\cite{HILLERY1984121,sch1,Leo1,CarlosNB15}. For a generic quantum state $\hat{\rho}$, it can be written as the expectation value of the displaced parity operator~\cite{Royer77}
\begin{equation}
W_{\hat{\rho}}(\bm{r}):=\mathrm{tr}\{\hat{\rho}\hat{D}(\alpha)\hat{\Pi}\hat{D}^\dagger(\alpha)\},
\end{equation}
where
\begin{equation}
\bm{r}:=(x,p)^\top:=2\left(\operatorname{Re}\{\alpha\},\operatorname{Im}\{\alpha\}\right)^\top,
\end{equation}
is the coordinate vector in phase space and $\hat{\Pi}:=(-1)^{\hat{a}^\dagger\hat{a}}$ is the parity operator. Using the well-known properties
\begin{subequations}\label{eq:displacements_parity_relations}
\begin{align}
\hat{D}(\alpha)\ket{\beta}=&\mathrm{e}^{\text{iIm}\{\alpha\beta^*\}}\ket{\alpha+\beta},\label{eq:prop1}
\\
\braket{\alpha|\beta}=&\text{e}^{-\text{i}\text{Im}\{\alpha\beta^*\}-\left|\alpha-\beta\right|/2},\label{eq:prop2}
\\
\hat{\Pi}\ket{\alpha}=&\ket{-\alpha},\label{eq:prop3}
\end{align}
\end{subequations}
it is easy to find for the arbitrary coherent-state superposition~(\ref{eq:coherent_superposition}) that
\begin{equation}\label{eq:coherent_superpositions}
W_{\ket{\psi}}(\bm{r})=\sum_{nm} \psi_n \psi^*_m W_{\ket{\alpha_n}\bra{\alpha_m}}(\bm{r}),
\end{equation}
where the Wigner function of the operator $\ket{\alpha_n}\bra{\alpha_m}$ is the complex Gaussian
\begin{align}
W_{\ket{\alpha_n}\bra{\alpha_m}}(\bm{r})=&\frac1{2\pi} \exp\bigg[-\left(\bm{r}-\frac{\bm{r}_n+\bm{r}_m}{2}\right)^2\bigg]
\\
&\times\exp\bigg[-\frac{\text{i}}2(\bm{r}_n-\bm{r}_m)^\top\Omega \bm{r}+\frac{\text{i}}{4}\bm{r}_n^\top\Omega \bm{r}_m\bigg],\nonumber
\end{align}
where
\begin{equation}
	\bm{r}_n:=2\left(\operatorname{Re}\{\alpha_n\},\operatorname{Im}\{\alpha_n\}\right)^\top,
\end{equation}
represents the locations of the coherent states in phase space and $\Omega:=\big(\begin{smallmatrix}
0 & 1\\
-1 & 0
\end{smallmatrix}\big)$ is known as the symplectic form.

The product of quadrature uncertainties has a lower limit $\Delta x\Delta p=1$, which is sometimes denoted by the \textit{Planck action} in phase space (note that it is equal to $\nicefrac{\hbar}{2}$ for position and momentum with proper units). This led to the belief that phase-space structures with areas below this Planck scale either do not exist or pose no observational consequences for physical quantum states. This was challenged by Zurek~\cite{Zurek2001}, who showed that compass states not only have \textit{sub-Planck} structures, but play a crucial role in enhancing their sensitivity to phase-space displacements, which is related to their metrological power as we  discuss later.

In order to discuss in more detail the origin of sub-Planck structures, we consider next some specific classes of coherent-state superpositions. Let us start with cat states~\cite{Gerry05book}, in particular the horizontal-cat defined along the position phase-space axis (in the following we omit normalizations of states and Wigner functions to simplify the notation)
\begin{equation}
\label{eq:horizontalcat_HW}
\ket{\psi_\text{H}}:=\ket{x_0/2}+\ket{-x_0/2},
\end{equation}
with $x_0 \in \mathbb{R}$. Particularizing the general Wigner function~(\ref{eq:coherent_superpositions}) to this state we obtain 
\begin{equation}
\label{eq:cat_Wigner_HW}
W_{\ket{\psi_\text{H}}}(\bm{r})
=\text{e}^{-\frac1{2}p^2}
\left[V(x;x_0)
+2\text{e}^{-\frac1{2}x^2}\cos\left( x_0 p\right)\right],
\end{equation}
for
\begin{equation}
\label{eq:doublepeak}
V(x;x_0)
:=\text{e}^{-\frac1{2}\left(x-x_0\right)^2}
+\text{e}^{-\frac1{2}\left(x+x_0\right)^2},
\end{equation}
which we plot in Fig.~\ref{fig:Wigner_HW}(a). The figure presents two Gaussian lobes centered at positions $(\pm x_0,0)$, which correspond to the Wigner functions of the underlying coherent states $\ket{\pm\nicefrac{x_0}{2}}$. In addition, around the origin of phase space, it shows oscillations along the momentum direction with period $\nicefrac{2\pi}{x_0}$, generated from the quantum interference between the coherent states. 

We can also build vertical cat states as a superposition of two coherent states in the momentum axis, and hence just as a $\pi/2$ rotation of the horizontal cat states, that is, 
\begin{equation}
	\ket{\psi_\text{V}}:=\exp\left(\frac{\text{i}\pi\hat{a}^\dagger\hat{a}}{2}\right)\ket{\psi_\text{H}}. 
\end{equation}
The corresponding Wigner function has the same form as Eq.~(\ref{eq:cat_Wigner_HW}), but swapping $x$ and $p$ as
\begin{equation}
	W_{\ket{\psi_\text{V}}}(x,p)=W_{\ket{\psi_\text{H}}}(p,x).
\end{equation}

Let us consider now the superposition of horizontal and vertical cat states, leading to Zurek's compass state~\cite{Zurek2001}
\begin{equation}
\label{eq:compass_HW}
\ket{\psi_{\text{C}}}:=\ket{\psi_\text{H}}+\ket{\psi_{\text{V}}}.
\end{equation}
The corresponding Wigner function is represented in Fig.~\ref{fig:Wigner_HW}(b), which using Eq.~(\ref{eq:coherent_superpositions}) can be written as
\begin{align}
W_{\ket{\psi_{\text{C}}}}(\bm{r})=W_{\text{coh}}(\bm{r})+2W_{\text{cent}}(\bm{r})+2W_\text{int}(\bm{r}),
\label{eq8}
\end{align}
where
\begin{equation}
W_{\text{coh}}(\bm{r})
:=\text{e}^{-\frac1{2}p^2}V(x;x_0)
+\text{e}^{-\frac1{2}x^2}V(p;x_0),
\end{equation}
corresponds to the Wigner functions of the four coherent states underlaying the compass state (Gaussian lobes at the north, south, east, and west positions),
\begin{equation}
	W_{\text{cent}}(\bm{r}):=\text{e}^{-\frac1{2}(p^2+x^2)}\big[\cos (x_0 p)+\cos (x_0 x)\big],
\end{equation}
corresponds to their interference pattern at the center of phase space, and
\begin{equation}
    W_{\text{int}}(\bm{r}):=\sum_{\sigma_x,\sigma_p=\pm 1}G(\sigma_x x,\sigma_p p),
\end{equation}
for
\begin{equation}
    G(x,p):=\text{e}^{-\frac1{2}\left[\left(x-\frac{x_0}{2}\right)^2+\left(p-\frac{x_0}{2}\right)^2\right]}\cos\Big[\frac{x_0}{2}\left(x+p-\frac{x_0}{2}\right)\Big],
\end{equation}
%
contains the interference terms generated far from the phase-space origin (cat-like interference patterns located at the northeast, northwest, southeast, and southwest positions).

Remarkably, the central chessboard-like pattern $W_{\text{cent}}(\bm{r})$ contains tiles of alternating sign (denoted by different colors in the figure) with areas proportional to $x_0^{-2}$, hence below the Planck scale for $x_0\gg 1$. These are the sub-Planck structures first identified in~\cite{Zurek2001}, whose size is limited in all phase space directions. In contrast, the structures appearing in the interference pattern of the cat state of Fig.~\ref{fig:Wigner_HW}(a) are limited only in the vertical direction, but not in the horizontal direction, where they show a Gaussian profile.

It is interesting for our purposes to consider one more type of states, built as incoherent mixtures of the cat states introduced above
\begin{align}
\label{eq:catmixture_HW}
\hat{\rho}_\text{M}:=\ket{\psi_{\text{H}}}\bra{\psi_{\text{H}}}+\ket{\psi_{\text{V}}}\bra{\psi_{\text{V}}}.
\end{align}
The corresponding Wigner function, shown in Fig.~\ref{fig:Wigner_HW}(c), reads
\begin{equation}
W_{\hat{\rho}_\text{M}}(\bm{r})
=W_{\text{coh}}(\bm{r})+2W_{\text{cent}}(\bm{r}),
\end{equation}
which is almost the same as that of the compass state, just missing the $W_{\text{int}}(\bm{r})$ interference terms. Remarkably, this means that the same chessboard pattern with sub-Planck structures appears for the cat mixture~\cite{Eff12}. Now, since the interference structures of cat states are not considered to be sub-Planck (since they are not limited in all phase-space directions), and an incoherent mixing of two states cannot enhance their individual quantum properties, this raises legitimate doubts about up to what point the presence of sub-Planck structures in the Wigner function is enough to claim that any interesting quantum properties appear.

\begin{figure*}[t]
\includegraphics[width=\textwidth]{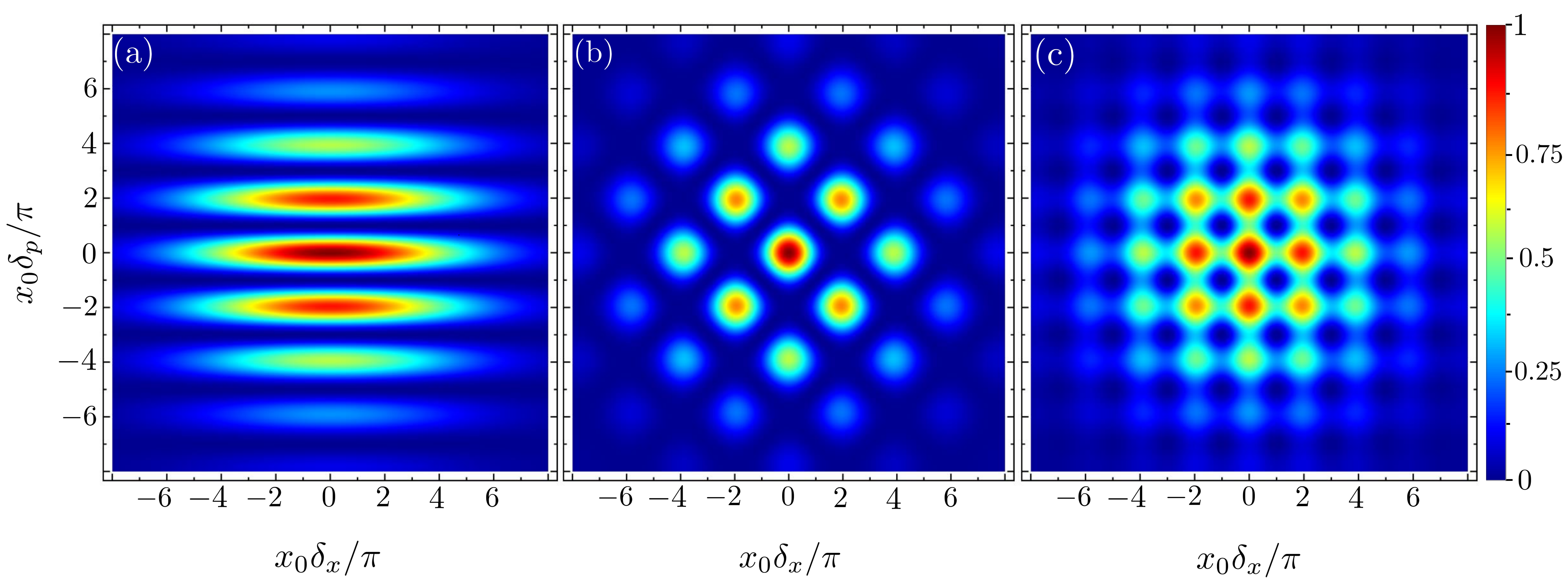}
\caption{Overlap between the (Heisenberg-Weyl) states considered in this work and their displaced versions, as a function of the real and imaginary parts of the displacement $\delta\alpha=\delta_x+\text{i}\delta_p$ normalized to $\pi/x0$: (a)~horizontal cat state, (b)~to the compass state, and (c)~to the cat-state mixture. In all cases $x_0=8$.}
\label{fig:overlap_HW}
\end{figure*}

\subsection{Sensitivity to displacements}\label{subsec:overlap_HW}

In order to show the differences between the three types of states introduced in the preceding section, it is interesting to consider their sensitivity to phase-space displacements. In particular, given a state $\hat{\rho}$ and its displaced version $\hat{D}(\delta\alpha)\hat{\rho}\hat{D}^\dagger(\delta\alpha)$, where $\delta\alpha\in\mathbb{C}$ is an arbitrary displacement, we consider their overlap
\begin{equation}
\label{eq:overlap_HW}
F_{\hat{\rho}}(\delta\alpha)
:=\text{tr}\{\hat{\rho}\hat{D}(\delta\alpha)\hat{\rho}\hat{D}^\dagger(\delta\alpha)\}	
=\left|\braket{\psi|\hat{D}(\delta\alpha)|\psi}\right|^2,
\end{equation}
where the last equality applies when the state is pure, $\hat{\rho}=\ket{\psi}\bra{\psi}$. This quantity provides a measure~\cite{Audenaert14} for the distinguishability of the state and its displaced version. The smaller the displacement $\delta\alpha$ needs to be in order to bring the overlap to zero, the more sensitive the state is said to be against displacements. This has immediate implications for quantum metrology: Imagine that a signal that we want to measure is linearly coupled to our harmonic oscillator, which then experiences a displacement proportional to the signal's strength; the sensitivity of the oscillator to displacements is then translated into a signal resolution, such that quantum states with higher sensitivity are able to resolve weaker signals~\cite{Toscano06}. 

Let us consider the example of coherent states. Using~(\ref{eq:displacements_parity_relations}), we easily find the overlap as 
\begin{equation}
	F_{\ket{\alpha}}\left(\delta \alpha\right)=\mathrm{e}^{-\left|\delta\alpha \right|^2}.
	\end{equation}
We see that displacements above the Planck scale $|\delta\alpha|>1$ are required in order to make this overlap vanish. This determines the resolution of coherent states as metrological tools. It is interesting to note that,
in the case of coherent states, the sensitivity to displacements is independent of the number of quanta present in the state, $N=\braket{\hat{a}^\dagger\hat{a}}=|\alpha|^2$. Hence, the sensitivity cannot be improved by increasing $N$ and is solely limited by the \textit{shot noise} introduced by vacuum fluctuations.

Consider next cat states, in particular the horizontal cat state defined in Eq.~(\ref{eq:horizontalcat_HW}).
Decomposing the displacement in real and imaginary parts as
\begin{equation}
\delta\alpha
=\frac{\delta_x+\text{i}\delta_p}{2},\hspace{3mm} \delta_j\in\mathbb{R},
\end{equation}
the overlap~(\ref{eq:overlap_HW}) is easily found to be
\begin{equation}\label{eq:overlap_horizontalcat_HW}
F_{\ket{\psi_\text{H}}}\left(\delta\alpha\right)=\frac1{2}\text{e}^{-|\delta\alpha|^2}\left[1+\cos(x_0 \delta_p)\right].
\end{equation}
Here and in the following we make heavy use of the properties~(\ref{eq:displacements_parity_relations}). The overlap function  $F_{\ket{\psi_\text{H}}}\left(\delta\alpha\right)$ vanishes identically when
 \begin{equation}
 	\delta_p=\frac{(2n+1)\pi}{x_0},\,n\in\mathbb{Z}.
 	\end{equation}
 
Note that the number of excitations in the cat state is $N\approx \nicefrac{x_0^2}{4}$ for $x_0\gg 1$. Hence, in contrast to coherent states, the sensitivity of cat states to displacements $\delta\alpha=\text{i}\delta_p$ in the momentum direction scales as $\nicefrac1{\sqrt{N}}$, which is well beyond the Planck scale for ``macroscopic'' cat states with $x_0\gg 1$ (this scaling is known as the \textit{Heisenberg limit} of the sensitivity). On the other hand, for displacements $\delta\alpha=\delta_x$ in the position direction, cat states pose no advantage with respect to coherent states. This is illustrated in Fig.~\ref{fig:overlap_HW}(a), where we plot the overlap as a function of the displacement.

Consider now the compass state of Eq.~(\ref{eq:compass_HW}), for which we easily obtain
\begin{align}
F_{\ket{\psi_\text{C}}}(\delta\alpha)=&\nonumber\frac{\mathrm{e}^{-|\delta\alpha|^2}}{4} \left[\cos\left(\frac{x_0 \delta_x}{2}\right)+\cos\left(\frac{x_0 \delta_p}{2}\right)\right]^2
\\
=&\frac{\mathrm{e}^{-\left|\delta\alpha\right|^2}}{2}\cos^2\left(\frac{x_0 \delta_+}{4}\right)\cos^2\left(\frac{x_0\delta_-}{4}\right),
\end{align}
with $\delta_\pm=\delta_x\pm\delta_p$. This overlap is shown in Fig.~\ref{fig:overlap_HW}(b). The vanishing condition is now either of the following
\begin{equation}
	\delta_x\pm\delta_p=\frac{2(2n+1)\pi}{x_0},~n\in\mathbb{Z}.
\end{equation}
 As appreciated in Fig.~\ref{fig:overlap_HW}(b), this is satisfied for displacements with $\left|\delta\alpha\right|\sim x_0^{-1}$ and arbitrary phase. Hence, as compared to coherent states, a compass state with $N$ excitations (approximately equal to $\nicefrac{x_0^2}{4}$ for $x_0\gg 1$) has a $\sqrt{N}$-enhanced sensitivity against displacements, but now in any phase-space direction, making it much more valuable for quantum metrology than cat states~\cite{Toscano06}.

There is a natural impulse to associate this property of compass states with their sign-alternating sub-Planck structures. However, only by themselves, these structures are not enough to generate such enhancement of the sensitivity to arbitrary displacements. In order to show this, consider now the cat-state mixture~(\ref{eq:catmixture_HW}), which as mentioned above also presents the same sub-Planck structures in the center of phase space. The overlap~(\ref{eq:overlap_HW}) reads in this case
\begin{equation}
F_{\hat{\rho}_\text{M}}(\delta\alpha)=\frac1{4}\mathrm{e}^{-|\delta\alpha|^2}\left[2+\cos(x_0\delta_x)+\cos(x_0\delta_p)\right].
\end{equation}
In order for this expression to vanish, the following two conditions need to be satisfied simultaneously:
\begin{subequations}
\begin{align}
\delta_x
&=\frac{(2n_x+1)\pi}{x_0},~n_x\in\mathbb{Z},
\\
\delta_p
&=\frac{(2n_p+1)\pi}{x_0},~n_p\in\mathbb{Z}.
\end{align}
\end{subequations}
Hence, the cat-state mixture also has $\sqrt{N}$-enhanced sensitivity against displacements, but only for those performed along the directions diagonal with respect to the axis system formed by the cat states ($\pm 45^\circ$ in our case). This is illustrated in Fig.~\ref{fig:overlap_HW}(c). Hence, in spite of also having sub-Planck structures in the Wigner function, cat-state mixture do not have the potential for metrology of compass states, for which the extra quantum coherence of the cat-state superposition plays a crucial role.

In summary, as compared to coherent states, for a given number of quanta $N$, compass states show a $\sqrt{N}$-enhancement of the sensitivity against displacements in arbitrary directions in phase space. In contrast, cat states and cat-state mixtures show this sensitivity enhancement only in specific phase-space directions. Interestingly, this shows that even though the same sub-Planck scales are present in the Wigner function of cat-state mixtures and superpositions, the latter have way more potential for quantum metrology.

\section{Generalization to the SU(2) group}\label{sec:SU(2)}

Quantum mechanics associates a Hilbert space to each physical system. In turn, one common way of characterizing Hilbert spaces is through an operator algebra. For example,
in~\S\ref{sec:HW} we considered the~$\mathfrak{hw}(1)$ algebra, which acts on an infinite-dimensional Hilbert space and is typically associated to one-dimensional mechanical systems. As different systems are characterized by different algebras, from the early days of quantum mechanics there was an interest in extending results found for one algebra to different ones, or, even better, developing full generalizations when possible~\cite{perelomov1986}. This is of even more practical interest these days, since we are now able to devise experimental implementations of essentially any algebra we can think of.

In this spirit, in this section we show that the previous results found for the~$\mathfrak{hw}(1)$ algebra can be extended to another of the most common algebras appearing in quantum-mechanical systems: the $\mathfrak{su}$(2) or angular momentum algebra. This algebra involves a vector operator $\hat{\bm{J}}:=(\hat{J_1},\hat{J_2},\hat{J_3})$, the generator of rotations~\cite{QuantumBook1,QuantumBook2,QuantumBook3}, and is characterized by the commutation relations
\begin{equation}
\left[\hat{J_n},\hat{J}_m\right]=\text{i} \sum^3_{l=1} \epsilon_{nml} \hat{J}_l,
\end{equation}
where $\epsilon_{nml}$ is the Levi-Civit\`{a} symbol (totally antisymmetric form). The irreducible representations of this algebra can be labeled by an index $j$ that can only take integer or half-integer values, and is related to the spectrum of $\hat{\bm{J}}^2$, the quadratic Casimir operator of the $\mathfrak{su}$(2) algebra, which commutes with all $\mathfrak{su}$(2) generators. Each of these representations has dimension $2j+1$, and is spanned by the common eigenbasis $\{\ket{j,\mu}\}_{\mu
=-j,-j+1,\ldots,j}$ of $\hat{\bm{J}}^2$ and one projection of $\hat{\bm{J}}$, say $\hat{J}_3$, so that
\begin{equation}
\hat{\bm{J}}^2\ket{j,\mu}=j\left(j+1\right)\ket{j,\mu},\hspace{2mm}\hat{J_3}\ket{j,\mu}=\mu\ket{j,\mu}.
\end{equation}

In the following we provide a SU(2) generalization of the various coherent-state superpositions discussed in~\S\ref{sec:HW}, showing that even their properties related to sub-Planck structures and sensitivity against displacements can be properly adapted to this case. In order to make a connection with the preceding section, it is interesting to note that, in quantum-optical experimental implementations, the maximal $\hat{J}_3$ eigenvalue $j$ is typically related to a number of excitations that need to be deposited in the system. For example, an $\mathfrak{su}$(2) algebra with $j$ angular momentum can be implemented with an ensemble of $2j$ indistinguishable two-level systems ($\ket{j,j}$ corresponding to all atoms excited) or with two bosonic modes sharing $2j$ excitations (this is the so-called Schwinger representation of angular momentum, where $\ket{j,j}$ corresponds to all excitations  gathered on one of the modes, leaving the other in vacuum)~\cite{MetrologyRMP18}. We  see that the $\sqrt{N}$-sensitivity enhancement of HW coherent-state superpositions is then replaced by a~$\sqrt{j}$ enhancement.

\subsection{Preliminaries:\\ SU(2) coherent states and the Wigner function}\label{subsec:preliminaries}

Let us introduce the SU(2) coherent states~\cite{Radcliffe1971,Perelomov72,GILMORE1972391,Arecchi72,perelomov1986}. Our starting point is the SU(2) displacement operator, which admits either of the forms
\begin{equation}\label{eq:SU(2)_displacement}
\hat{D}(\gamma):=\mathrm{e}^{\alpha\hat{J}_--\alpha^*\hat{J}_+}=\mathrm{e}^{\gamma\hat{J}_-}\text{e}^{-\eta\hat{J}_3}\text{e}^{-\gamma^*\hat{J}_+}=\mathrm{e}^{\text{i}\theta \bm{n}\cdot\hat{\bm{J}}},
\end{equation}
parametrized through $\gamma=\text{e}^{i\phi}\tan(\theta/2)$ for convenience, with $\phi\in[0,2\pi[$ and $\theta\in[0,\pi]$, in which case $\alpha=\text{e}^{\text{i}\phi}\theta/2$, $\eta=\ln(1+|\gamma|^2)$, and $\bm{n}=(\sin\phi,-\cos\phi,0)$. The raising and lowering operators are
\begin{equation}
	\hat{J}_\pm:=\hat{J}_1\pm \text{i} \hat{J}_2.
\end{equation}
The SU(2) displacements correspond to rotations of angle $-\theta$ around the $\bm{n}$ axis. Throughout the article, we make use of their composition rule~\cite{perelomov1986}
\begin{equation}\label{eq:composition_rule}
\hat{D}(\gamma_1)\hat{D}(\gamma_2)=\hat{D}(\gamma_3)\text{e}^{{\text{i}}\varphi\hat{J_3}},
\end{equation}
where
\begin{subequations}
\begin{align}
	\gamma_3=&\frac{\gamma_1+\gamma_2}{1-\gamma^*_1\gamma_2},
\\ \varphi=&2\text{arg}(1-\gamma^*_1\gamma_2).
\end{align}
\end{subequations}
As Eq.~(\ref{eq:composition_rule}) does not depend on $j$, it is easily proven by particularizing it to the $j=\nicefrac1{2}$ representation, where all operators correspond to $2\times 2$ matrices (Pauli matrices for the components of $\hat{\bm{J}}$).

Applying the displacement operator to the reference state $\ket{j,j}$, we obtain a SU(2) coherent state
\begin{equation}\label{eq:SU(2)_coherent}
\ket{\gamma}:=\hat{D}(\gamma)\ket{j,j}=\sum^j_{\mu=-j} u_\mu(\gamma)\ket{j,\mu},
\end{equation}
where
\begin{equation}
u_\mu(\gamma)
=\sqrt{\frac{(2j)!}{(j+\mu)!(j-\mu)!}}\frac{\gamma^{j-\mu}}{(1+|\gamma|^2)^j},
\end{equation}
is easily found by using the intermediate form of the displacement~(\ref{eq:SU(2)_displacement}), taking into account that
\begin{subequations}
	\begin{align}
		\hat{J}_+\ket{j,j}=&0,\\ \hat{J}_-\ket{j,\mu}=&N_\mu\ket{j,\mu-1},
	\end{align}
\end{subequations}
where $N_\mu=\sqrt{(j+\mu)(j-\mu+1)}$. The coherent states can be associated with the points 
\begin{equation}
\left(\sin\theta\cos\phi,\sin\theta\sin\phi,\cos\theta\right)
\end{equation}
on a unit sphere, such that $\ket{j,j}$ corresponds to the north pole. 
The overlap between two coherent states is
\begin{equation}\label{eq:overlap_SU(2)}
\braket{\gamma_1|\gamma_2}=\left[\frac{\left(1+\gamma^*_1 \gamma_2\right)^2}{(1+|\gamma_1|^2)(1+|\gamma_2|^2)}\right]^j,
\end{equation}
as easily proven from Eqs.~(\ref{eq:composition_rule}) and~(\ref{eq:SU(2)_coherent}).

Hence, for $j\gg 1$ the overlap between the reference state $\ket{0}=\ket{j,j}$ and any coherent state $\ket{\gamma}$ is approximately Gaussian as a function of $\gamma$,
\begin{align}\label{eq:overlap_SU(2)_Planckscale}
\braket{0|\gamma}=\frac1{(1+|\gamma|^2)^j}\approx \mathrm{e}^{-j|\gamma|^2},
\end{align}
similarly to the HW coherent states. Note that with this parametrization of SU(2) coherent states in terms of $\gamma$, the width of the Gaussian scales as $\nicefrac{1}{\sqrt{j}}$. In general, the overlap between two coherent states pinned at different points of the unit sphere decreases with $j$ as
\begin{align}
\left|\braket{\gamma_1|\gamma_2}\right|=\cos^{2j}(\Theta/2),
\end{align}
where
\begin{equation}
\cos\Theta=\cos\theta_1\cos\theta_2+\sin\theta_1\sin\theta_2\cos(\phi_1-\phi_2),
\end{equation}
for
\begin{equation}
\gamma_n=\text{e}^{\text{i}\phi_n}\tan(\theta_n/2).
\end{equation}
Indeed, we  see later that the effective support of the SU(2) coherent-state's Wigner function (which can be defined as a quasi-probability distribution on the sphere) decreases following the corresponding scaling in $j$.

\begin{figure*}[t]
\includegraphics[width=\textwidth]{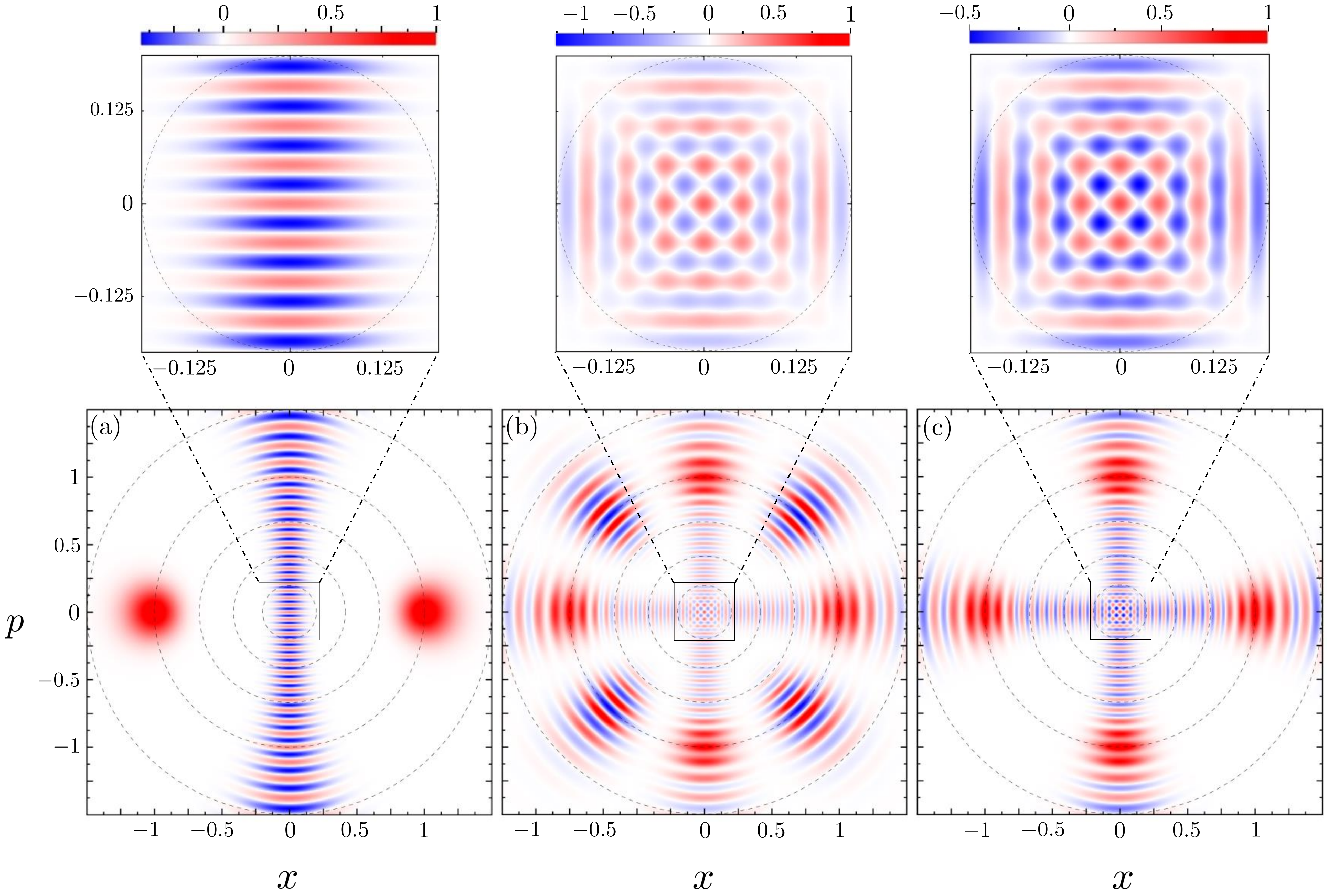}
\caption{Stereographic projection of the SU(2) Wigner functions of the coherent-state superpositions considered in this work: (a)~horizontal cat state, (b)~compass state, and (c)~cat-state mixture. In all cases $j=30$. Red and blue regions correspond, respectively, to positive and negative values of the function, which we normalize to its maximum value. The concentric grey dashed circles represent great circles in the sphere with polar angles $\{n\pi/8\}_{n=1,2,\ldots,5}$, such that $n=4$ corresponds to the equator.}
\label{fig:Wigner_SU(2)}
\end{figure*}

In the next section we consider different coherent-state superpositions of the type
\begin{equation}\label{eq:SU(2)_superposition}
\ket{\psi}=\sum_n \psi_n \ket{\gamma_n},~\psi_n\in\mathbb{C},
\end{equation}
which  generalize the ones studied for the~$\mathfrak{hw}(1)$ algebra. Just as in that case, the Wigner function is a very useful tool for the characterization of these states~\cite{Varilly89,Heiss00,Klimov17,Glaser20}. In the SU(2) case, for any state $\hat{\rho}$, it can be evaluated via the Stratonovich-Weyl correspondence as~\cite{davis2020},
\begin{equation}
W_{\hat{\rho}}(\gamma):=\mathrm{tr}\{\hat{\rho}\hat{D}(\gamma)\hat{\Delta}\hat{D}^\dagger(\gamma)\},
\end{equation}
with
\begin{equation}
\hat{\Delta}:=\sum^j_{\mu=-j}\Delta_\mu\ket{j,\mu}\bra{j,\mu},
\end{equation}
where
\begin{equation}
\label{eq:Deltamu}
\Delta_\mu
:=\sum^{2j}_{l=0}\frac{2l+1}{2j+1}\braket{j,\mu;l,0|j,\mu},
\end{equation}
with~$\braket{j,\mu;l, 0|j,\mu}$ the Clebsch-Gordan coefficients. As for the Heisenberg-Weyl group, this Wigner function is real and bounded (but not necessarily positive, not even for coherent states in this case, although their negativity decreases with $j$~\cite{davis2020}). Hence, we can use it to visualize the state, by plotting it either over the surface of the unit sphere or, as we  do, in the stereographic plane $(x,p)$, where
\begin{equation}
\gamma:=x+\text{i}p.
\end{equation}
Note that the origin of this plane corresponds to the north pole of the sphere, while the equator is represented as the circle of unit radius, and the south pole as a circle of infinite radius.

The Wigner function of the generic superposition~(\ref{eq:SU(2)_superposition}) can be written as
\begin{equation}
W_{\ket{\psi}}(\gamma)=\sum_{nm} \psi_n \psi^*_m W_{\ket{\gamma_n}\bra{\gamma_m}}(\gamma),
\end{equation}
where, using Eqs.~(\ref{eq:composition_rule}) and~(\ref{eq:SU(2)_coherent}), the Wigner function of the operator $\ket{\gamma_n}\bra{\gamma_m}$ is easily found to be
\begin{equation}\label{eq:SU(2)_superposition_Wigner}
W_{\ket{\gamma_n}\bra{\gamma_m}}(\gamma)=\text{e}^{\text{i}\left(\varphi_n-\varphi_m\right)}\sum^j_{\mu=-j}\Delta_\mu u^*_\mu\left(\gamma^\prime_{m}\right)u_\mu\left(\gamma^\prime_{n}\right),
\end{equation}
with
\begin{subequations}
\begin{align}
	\varphi_k=&2\text{arg}(1+\gamma^*\gamma_k),
\\
\gamma_k^\prime=&\frac{\gamma_k-\gamma}{1+\gamma^*\gamma_k}.
\end{align}
\end{subequations}
This form is suited for numerical computation,
but, for analytical purposes, sometimes alternative derivations are more convenient, as we show in the Appendix for the states that we consider in the next section.
\subsection{SU(2) cat and compass states: \\sub-Planck structures}\label{subsec:sub-Planck_structures_SU(2)}

Let us now consider specific coherent-state superpositions, and show how the concept of sub-Planck structures presented in~\S\ref{sec:HW} for the~$\mathfrak{hw}(1)$ algebra, also extends to the $\mathfrak{su}(2)$ case. To this aim, we focus on coherent states distributed along the equator, that is, $\left|\gamma_n\right|=1$ in Eq.~(\ref{eq:SU(2)_superposition}).

Superposing two antipodal SU(2) coherent states we obtain a SU(2) cat state~\cite{Sanders89,Huang15,Huang18,Maleki,davis2020}. Consider in particular the coherent states along the horizontal axis of the stereographic plane (as in~\S\ref{sec:HW}, we omit normalizations in states and Wigner functions), defining the horizontal cat state
\begin{equation}\label{eq:horizontalcat_SU(2)}
\ket{\psi_\text{H}}:=\ket1+\ket{-1}.
\end{equation}
We derive the Wigner function of this state in the Appendix, obtaining
\begin{equation}
\label{eq:horizontalcat_SU(2)_Wigner}
W_{\ket{\psi_\text{H}}}(\gamma)= W_{\ket1}(\gamma)+W_{\ket{-1}}(\gamma)+2I_\text{H}(\gamma),
\end{equation}
where the last term provides the interference between the underlying coherent states
\begin{equation}\label{eq:horizontalcat_SU(2)_interference}
I_\text{H}(\gamma)
:=\sqrt{\frac{\left(4j+1\right)!}{2j+1}}\frac{\cos(2j\bar\phi)\sin^{2j}\bar\theta}{4^j (2j)!},
\end{equation}
with
\begin{subequations}
\begin{align}
\bar\phi &=\arg\{\gamma-1\}+\arg\{\gamma+1\},
 \\
 \bar\theta &=2\text{tan}^{-1}\left|\frac{\gamma-1}{\gamma+1}\right|,	
\end{align}
\end{subequations}
while the first terms correspond to the Wigner functions of the coherent states
\begin{equation}\label{eq:coherent_terms_Wigner}
W_{\ket{\pm1}}(\gamma)=\frac{(2j)!}{\sqrt{2j+1}}\sum^{2j}_{l=0}\frac{(2l+1)P_l(\pm\cos\bar\theta)}{\sqrt{(2j-l)!(2j+l+1)!}},
\end{equation}
where $P_l(x)$ is the $l\mathrm{th}$ Legendre polynomial~\cite{Varilly89}.

In Fig.~\ref{fig:Wigner_SU(2)}(a)~we plot this Wigner function in the stereographic plane $(x,p)$, as explained in the preceding section. We can clearly see two lobes centered at positions $(\pm 1,0)$, which correspond to the coherent states. Increasing $j$ has the effect of reducing the extension (effective support) of these lobes. In addition, the interference between these coherent states generates an oscillating pattern in the vertical direction, similarly to what happens for the horizontal HW cat states. The zeros of this interference pattern $I_\text{H}(\gamma)$, occur when 
\begin{align}
	2j\bar\phi=\frac{(2n+1)\pi}{2},~n\in\mathbb{Z}.
\end{align}
Along the $p$ axis ($x=0$), this means that the first zeros are located at $p=\pm\tan(\pi/8j)\approx\pm\pi/8j$ for $j\gg 1$. On the other hand, in the horizontal direction for any fixed $p$, the interference pattern simply decays as a Gaussian $I_\text{H}(x)/I_\text{H}(0)\approx \text{e}^{-4jx^2}$ for $j\gg 1$, hence with a width proportional to $\nicefrac1{\sqrt{j}}$. This is precisely the same scaling that we found for the width of the overlap between coherent states. Hence, similarly to the HW case, we see that the support of the structures appearing in the interference pattern of horizontal SU(2) cat states is limited only in the $p$ direction.

We can also define cat states along the vertical axis of the stereographic plane as
\begin{align}
\label{eq:verticalcat_SU(2)}
\ket{\psi_\text{V}}
:=\ket{\text{i}}+\ket{-\text{i}},
\end{align}
whose Wigner function is like the one of the horizontal cat state, but rotated by $\pi/2$ in the stereographic plane, that is,
\begin{align}
	W_{\ket{\psi_\text{V}}}(\gamma)=&\nonumber W_{\ket{\text{i}}}(\gamma)+W_{\ket{-\text{i}}}(\gamma)+2I_\text{V}(\gamma)
	\\
	=& W_{\ket{\psi_\text{H}}}(p+\text{i}x).
	\end{align}
Note that, similarly to Eq.~(\ref{eq:horizontalcat_SU(2)_Wigner}) for $\ket{\psi_\text{H}}$, we write the Wigner function of $\ket{\psi_\text{V}}$ as the sum of the Wigner function of its underlaying coherent states $W_{\ket{\pm\text{i}}}(\gamma)=W_{\ket{\pm 1}}(p+\text{i}x)$, plus their interference $I_\text{V}(\gamma)=I_\text{H}(p+\text{i}x)$.

Given these cat states, we then define the SU(2) compass states as we already did for the HW case: the balanced superposition defined by
\begin{align}
	\ket{\psi_\text{C}}:=\ket{\psi_\text{H}}+\ket{\psi_\text{V}}.
\end{align}
The corresponding Wigner function, which we show in Fig.~\ref{fig:Wigner_SU(2)}(b), is equal to the sum of the individual Wigner functions of each cat state, plus the terms coming from the interference between these (the northwest, northeast, southeast, and southwest structures shown in the figure). We discuss the analytic form of this Wigner function in the Appendix. For our purposes here, it is interesting to note that for $j\gg 1$ the interference pattern close to the origin of the stereographic plane (say $|\gamma|<1$) is the sum of the interference patterns of the cat states, that is,
\begin{align}
W_{\text{cent}}(\gamma)=I_\text{H}(\gamma)+I_\text{V}(\gamma),
	\end{align}
leading to the chessboardlike pattern shown in Fig.~\ref{fig:Wigner_SU(2)}(b). This pattern is reminiscent of the one found for the~$\mathfrak{hw}(1)$ algebra, and shows that structures limited in all directions of the stereographic plane appear in this SU(2) compass state.
This alternating-sign tiles have an extension proportional
to $\nicefrac1{j}$ in any direction,
which is a factor $\nicefrac1{\sqrt{j}}$ smaller than the extension found for coherent states.
This generalizes the concept of sub-Planck structures to SU(2):
structures with a support of area $\nicefrac1{j}$ smaller than the support of coherent states.

From the previous discussion, it is clear that a mixture of cat states  also present the same chessboard pattern. Indeed, considering the mixture,
\begin{align}\label{eq:catmixture_SU(2)}
	\hat{\rho}_\text{M}:=\ket{\psi_\text{H}}\bra{\psi_\text{H}}+\ket{\psi_\text{V}}\bra{\psi_\text{V}},
	\end{align}
the corresponding Wigner function of above cat-state mixture can be simply written as,
 \begin{align}
 	W_{\hat{\rho}_\text{M}}(\gamma)=W_{\ket{\psi_\text{H}}}(\gamma)+W_{\ket{\psi_\text{V}}}(\gamma),
 	\end{align}
which is shown in Fig.~\ref{fig:Wigner_SU(2)}(c), where the chessboard can be appreciated. In fact, just like in the case of the~$\mathfrak{hw}(1)$ algebra, only the inter-cat-state interference regions are missing with respect to the Wigner function of the compass state.

In summary, we have shown that the same phase-space features found for HW cat and compass states, as well as cat-state mixtures, are found for their SU(2) counterparts in the stereographic plane, provided we restrict the underlying coherent states to the equator. Interestingly, the role of $x_0$ in the HW case (distance of the coherent states from the origin) is played by~$\sqrt{j}$ in the SU(2) case. Let us remark that, around the equator, the SU(2) group is contracted in the $j\rightarrow\infty$ limit to the two-dimensional Euclidean group E(2) group~\cite{barry2000}, which makes the results that we are exposing in this section nontrivial. In contrast, around the poles SU(2) is contracted to the HW group (Holstein-Primakoff approximation)~\cite{Arecchi72,barry2000}, in which case all the results of the preceding section are trivially generalized, since $\lim_{j\rightarrow\infty}
\nicefrac{\hat{J}_-}{\sqrt{j}}$
simply becomes an annihilation operator.

\begin{figure*}[t]
\includegraphics[width=\textwidth]{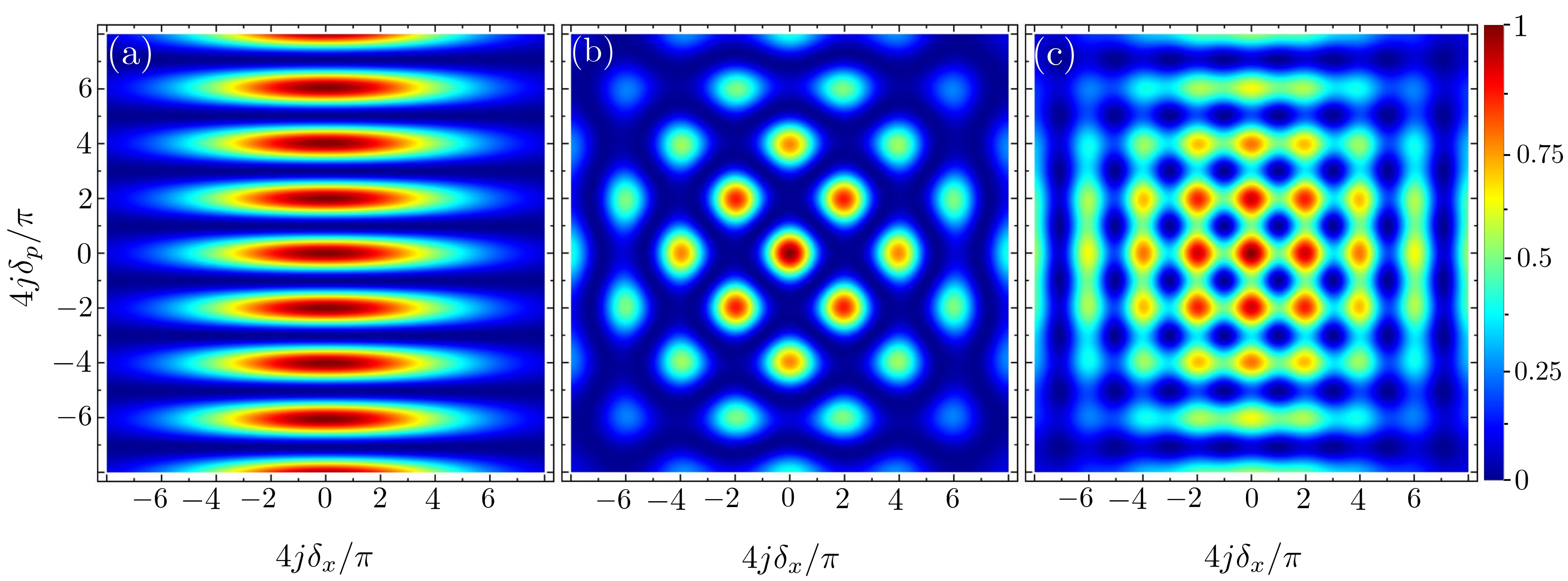}
\caption{Overlap between the SU(2) states considered in this work and their displaced versions, as a function of the real and imaginary parts of the displacement $\delta\gamma=\delta_x+\text{i}\delta_p$ normalized to $\pi/4j$. (a)~horizontal cat state, (b)~compass state, and (c)~cat-state mixture. In all cases $j=10$.}
\label{fig:overlap_SU(2)}
\end{figure*}

\subsection{Distinguishability under SU(2) displacements}\label{subsec:overlap_SU(2)}

Let us now show that the SU(2) coherent-state superpositions discussed above also behave similarly to their HW counterparts regarding their sensitivity against displacements. For this, again we compute the overlap between the states and their $\delta\gamma$-displaced versions, as given by Eq.~(\ref{eq:overlap_HW}). This overlap is easily evaluated for coherent-state superpositions by making use of the property
\begin{equation}\label{eq:Overlap_SU(2)_general}
\bra{\gamma_1}\hspace{-0.8mm}\hat{D}(\gamma)\hspace{-0.8mm}\ket{\gamma_2}=\left[\frac{(1-\gamma^*\gamma_1+\gamma\gamma_2^*+\gamma_1\gamma_2^*)^2}{2(1\hspace{-0.8mm}+\hspace{-0.8mm}|\gamma|^2)(1\hspace{-0.8mm}+\hspace{-0.8mm}|\gamma_1|^2)(1\hspace{-0.8mm}+\hspace{-0.8mm}|\gamma_2|^2)}\right]^j,
\end{equation}
which is proven by using the composition rule~(\ref{eq:composition_rule}) and the coherent-state overlap~(\ref{eq:overlap_SU(2)}).

In the case of a single coherent state, we have already established in~\S\ref{sec:HW} that, for $j\gg 1$, this overlap has an approximately Gaussian form $\text{e}^{-j|\delta\gamma|^2}$. Hence, in the SU(2) case, the sensitivity of coherent states to displacements scales as $\nicefrac1{\sqrt{j}}$. We then have to compare the sensitivity of coherent-state superpositions against this scaling.

Let us start with cat states. Defining $\delta\gamma:=\delta_x+\text{i}\delta_p$, with $(\delta_x,\delta_p)\in\mathbb{R}^2$, using Eq.~(\ref{eq:Overlap_SU(2)_general}), for the horizontal cat~(\ref{eq:horizontalcat_SU(2)}) we easily obtain
\begin{align}\label{eq:overlap_horizontalcat_SU(2)}
F_{\ket{\psi_\text{H}}}(\delta\gamma)
=\frac{\left[\delta_x^{2j}+(1+\delta_p^2)^j\cos(2j\text{tan}^{-1}\delta_p)\right]^2}{(1+|\delta\gamma|^2)^{2j}}.
\end{align}
We show this overlap in Fig.~\ref{fig:overlap_SU(2)}(a). As long as $j\gg 1$ and $\delta\gamma<1$, Eq.~(\ref{eq:overlap_horizontalcat_SU(2)}) shows that $\delta_x$ plays no relevant role in the overlap, which vanishes for displacements
\begin{align}
\delta_p=\tan\left[\frac{(2n+1)\pi}{4j}\right],~n\in\mathbb{Z}.
\end{align}
For large $j$, minimum displacement that turns the cat state into an orthogonal state scales as $\left|\delta\gamma\right|\sim \nicefrac1{j}$ and must occur in the vertical direction of the stereographic plane. Therefore, in this direction, as compared to coherent states, cat states show a $\sqrt{j}$-enhanced sensitivity to displacements. In contrast, for horizontal displacements they show no enhancement of the sensitivity compared to coherent states.

Let us now consider the compass state, for which the overlap~(\ref{eq:Overlap_SU(2)_general}) leads to
\begin{align}
F_{\ket{\psi_\text{C}}}(\delta\gamma)=\frac{\sum\limits_{q=x,p}\hspace{-1mm}\big[\delta_q^{2j}\hspace{-0.5mm}+(1+\delta_q^2)^j\cos(2j\text{tan}^{-1}\delta_q)\big]}{4(1+|\delta\gamma|^2)^{2j}}.
\end{align}
We show this overlap in Fig.~\ref{fig:overlap_SU(2)}(b). Similarly to the HW compass state, now the $\sqrt{j}$-enhanced sensitivity to displacements is independent of the displacement direction.

Finally, we calculate the overlap for the SU(2) cat-state mixture~(\ref{eq:catmixture_SU(2)}). The key point here is to note that, for $j\gg 1$ and $\left|\delta\gamma \right|<1$, the contribution of the cross terms between the cat states to the overlap, e.g., $\bra{\psi_\text{H}}\hat{D}(\delta\gamma)\ket{\psi_\text{V}}$, is negligible. Then, for small displacements the overlap is the sum of overlaps of the individual cat states, that is,
\begin{align}\label{eq:mixturecats_overlap_SU(2)}
F_{\hat{\rho}_\text{M}}(\delta\gamma)=F_{\ket{\psi_\text{H}}}(\delta\gamma)+F_{\ket{\psi_\text{V}}}(\delta\gamma),\text{ for }|\delta\gamma|<1,
\end{align}
where $F_{\ket{\psi_\text{V}}}(\delta\gamma)=F_{\ket{\psi_\text{H}}}(\delta_p+\text{i}\delta_x)$. The corresponding overlap is shown in Fig.~\ref{fig:overlap_SU(2)}(c), and shows the same properties as its HW counterpart: the $\sqrt{j}$-enhanced sensitivity is only present for displacements in the $\delta_p=\pm\delta_x$ directions.

These results show that, as promised, all the features we found for HW coherent-state superpositions are exported to SU(2) coherent-state superpositions, as long as the states are restricted to the equator, where the SU(2) group contracts to the Euclidean group in two dimensions E(2)~\cite{barry2000}, as already mentioned above. In particular, we have shown that SU(2) compass states have a $\sqrt{j}$-enhanced sensitivity against displacements as compared to coherent states. This occurs for displacements in arbitrary directions in the stereographic plane, while cat states and mixtures show such an enhanced sensitivity only in special directions.

Note that one of the main motivations for the study of SU(2) compass states is their potential application to quantum metrology, in particular as sensors of rotations. When the rotation axis is known, cat states along that axis are known to provide the desired quantum advantage over coherent states~\cite{Maccone2004,MetrologyDowling08,Maccone2011,MetrologyRMP18}. In contrast, when the axis of rotation is completely unknown, it has been found that the optimal quantum \textit{rotosensor states} are complicated in general \cite{Soto2015,Soto15,Coronado2017,Daniel2018,Martin2020}, but well approximated for small angles by anticoherent states, whose so-called \textit{Majorana constellation} is uniformly spread over the unit sphere \cite{Giraud2010}. This has been checked so far up to $j=5$, but it is believed to be true for arbitrary $j$ based on physical arguments \cite{Coronado2017,Martin2020}. Our compass states can be very useful for intermediate situations in which the rotation axis is bound to a given plane, but is otherwise unknown. This is because the SU(2) displacements defined in Eq.~(\ref{eq:SU(2)_displacement}) effect rotations around any axis defined on the equator of the unit sphere so that by orienting the sphere appropriately, compass states will provide $\sqrt{j}$-enhanced sensitivity to the rotations we want to measure as compared to coherent states.

\section{Conclusions and outlook}\label{sec:conclusions}

In this work we have generalized to the SU(2) group some features of coherent-state superpositions commonly studied for the HW group associated with the harmonic oscillator. Specifically, by restricting the SU(2) coherent states to the equator of the sphere that plays the role of phase space, we have shown that four-state superpositions have a Wigner function with properties similar to that of the HW compass states: A chess-board like pattern appears around the origin of the stereographic plane, which contains structures with support that scales as $\nicefrac1{j}$ with respect to the effective support of coherent states. This generalizes the sub-Planck structures found in HW compass states, with the role of the number of quanta $N$ being played by the total angular momentum $j$. Further, SU(2) cat states can also be defined as superpositions of two coherent states, which show an interference pattern with structures limited only along one direction, just as their HW counterparts. Moreover, SU(2) cat-state mixtures present the same sub-Planck structures as compass states. However, we have shown that compass states and cat-state mixtures are very different regarding their sensitivity to SU(2) displacements: As compared to coherent states, compass states have a $\sqrt{j}$-enhanced sensitivity against displacements in any direction, while cat states and cat-state mixtures show such enhancement only along specific directions. Again, this is exactly the same behavior found for the HW case, with $j$ playing the role of~$N$.

For $j\rightarrow\infty$ there is a well known connection between the HW group and the restriction of SU(2) to the neighborhood of a pole (Holstein-Primakoff approximation)~\cite{Arecchi72}. In contrast, for states around the equator, SU(2) contracts to the two-dimensional Euclidean group E(2)~\cite{barry2000}, which makes our generalizations nontrivial. Looking ahead, it is interesting to extend the notion of sub-Planck structures and the corresponding sensitivity to displacements to arbitrary groups, especially those relevant to modern experimental quantum-optical platforms such as SU(1,1)~\cite{perelomov1986,Seyfarth20} and higher-dimensional Lie groups including SU(3)~\cite{Nemoto2001}.

Having established these connections between the HW and SU(2) groups, there are many other routes that one can pursue. For example, in this work we have considered only superpositions of coherent states with the same amplitude. Interestingly, in the case of the HW group, superpositions of states with different amplitudes have been shown to play a role as eigenstates of the displaced-parity operator $\hat{B}(\alpha)=\hat{D}(\alpha)\hat{\Pi}$, which in turn is related to a nontrivial symmetry of the forced Harmonic oscillator \cite{Markovich2020}. Generalizing to the SU(2) case the parity operator as $\hat{\Pi}=(-1)^{j-\hat{J}_3}$, it is easy to show that the coherent-state superpositions $\ket{0}\pm\ket{\gamma}$ are eigenstates of the SU(2) displaced-parity operator $\hat{B}(\gamma)=\hat{D}(\gamma)\hat{\Pi}$, similarly to the HW case. It will then be interesting to extend to the SU(2) group all the results known in this context for the HW group, in particular exploring whether the latter operator is also related to a nontrivial symmetry of some SU(2) Hamiltonian. Moreover, generalized displaced-parity operators with very interesting properties have been defined for the HW group \cite{messina2013}, whose SU(2) counterparts can also be investigated in future works.

Another clear future endeavor would concern how to generate the SU(2) compass states introduced in our work. There is already a vast literature discussing plausible schemes for the implementation of SU(2) cat states, particularly NOON and GHZ states, as mentioned in the introduction (see \cite{Omran19,Monz11,Wang16,Afek10,Zhang16} for actual recent experimental implementations in different platforms). Perhaps some of these methods can be adapted for the generation of a superposition of the four coherent states required for compass states, which otherwise will require developing completely new proposals. As a concrete promising candidate, let us mention weak-field homodyne detection \cite{Walmsley2020}, which so far has been used to engineer HW cat states of arbitrary amplitude \cite{Thekkadath2020engineering}, but can be generalized for the robust preparation of compass states and should be exportable to some of the SU(2) platforms in which homodyne detection is available.

\begin{acknowledgements}
N.A. thanks everyone at the Wilczek Quantum Center for their hospitality and the Chinese Scholarship Council for financial support. B.C.S. acknowledges the National Natural Science Foundation of China  for their financial support (Grant No.~11675164). C.N.B. appreciates support from a Shanghai talent program and from the Shanghai Municipal Science and Technology Major Project (Grant No.~2019SHZDZX01).
\end{acknowledgements}

\appendix

\section{Wigner function of SU(2) states}\label{appendix:appendixA}

Let us here provide more detailed derivations of the various SU(2) Wigner functions that we have introduced in the main text. Instead of directly using the general form of Eq.~(\ref{eq:SU(2)_superposition_Wigner}), in some cases simpler expressions are found using the composition property of displacement operators, which we rewrite here as
\begin{equation}\label{appendix:composition_rule}
\hat{D}(\gamma_1)\hat{D}(\gamma_2)=\text{e}^{-\text{i}\varphi\hat{J_3}}\hat{D}(\gamma_3),
\end{equation}
with
\begin{subequations}
	\begin{align}
	\gamma_3=&\frac{\gamma_1+\gamma_2}{1-\gamma_1\gamma_2^*},\\ \varphi=&2\text{arg}(1-\gamma_1\gamma_2^*),
	\end{align}
\end{subequations}
and their action on the kernel $\hat{\Delta}$~\cite{Varilly89,Klimov17}:
\begin{align}\label{appendix:kernel}
\hat{D}(\gamma)\hat{\Delta}\hat{D}^\dagger(\gamma)=\sqrt{\frac{4\pi}{2j+1}}\sum^{2j}_{l=0}\sum^l_{m=-l}Y^*_{lm}(\theta,\phi)\hat{T}_{lm},
\end{align}
where $\gamma=\text{e}^{\text{i}\phi}\tan{\theta/2}$, and we have defined the spherical harmonics $Y_{lm}(\theta,\phi)$ and the irreducible tensor operators
\begin{equation}
\hat{T}_{lm}
:=\sqrt{\frac{2l+1}{2j+1}}\sum^j_{\mu,\mu^\prime=-j}\hspace{-2mm}\braket{j,\mu;l,m|j,\mu^\prime}\ket{j,n^\prime}\bra{j,n},
\end{equation}
where we recall that $\braket{j,\mu;l,m|j,\mu^\prime}$ are Clebsch-Gordan coefficients. In the following, we use these expressions to simplify the derivations of some Wigner functions.

Let us start with the Wigner function of the coherent state~$\ket1$, which is expressed as
\begin{equation}
W_{\ket1}(\gamma)=\braket{1|\hat{D}(\gamma)\hat{\Delta}\hat{D}^\dagger(\gamma)|1}.
\end{equation}
Using the composition rule~(\ref{appendix:composition_rule}) we can rewrite this expression as
\begin{align}
W_{\ket1}(\gamma)&=\braket{ j,j|\hat{D}(-1)\hat{D}(\gamma)\hat{\Delta}\hat{D}^\dagger(\gamma)\hat{D}(1)|j,j}
\\
&=\braket{ j,j|\hat{D}(\bar{\gamma})\hat{\Delta}\hat{D}^\dagger(\bar{\gamma})|j,j}
\end{align}
with $\bar\gamma=\nicefrac{(\gamma-1)}{(\gamma^*+1)}\equiv\text{e}^{\text{i}\bar\phi}\tan(\bar\theta/2)$. Inserting~(\ref{appendix:kernel}) in this expression, and taking into account that
\begin{subequations}\label{appendix:simplify_1}
\begin{align}
\braket{j,j;l,m|j,j}&=\delta_{m0}\sqrt{\frac{(2j)!^2(2j+1)}{(2j-l)!(2j+l+1)!}},
\\
Y_{l0}(\bar\theta,\bar\phi)&=\sqrt{\frac{2l+1}{4\pi}}P_l(\cos\bar\theta),
\end{align}
\end{subequations}
we obtain the expression~(\ref{eq:coherent_terms_Wigner}).

The Wigner function of the coherent state~$\ket{-1}$, which reads
\begin{equation}\label{WignerAux}
W_{\ket{-1}}(\gamma)=\bra{-1}\hat{D}(\gamma)\hat{\Delta}\hat{D}^\dagger(\gamma)\ket{-1},
\end{equation}
 is found in a similar fashion, but first we need to manipulate a bit the expression. In particular, we use the identity
\begin{align}
\ket{j,\mp j}&=\lim_{x\rightarrow \pm \infty}\hat{D}(x)\ket{j,\pm j}
\end{align}
to write
\begin{align}
\ket{-1}=\lim_{x\rightarrow-\infty}\hat{D}(-1)\hat{D}(x)\ket{j,-j}=\hat{D}(1)\ket{j,-j},
\end{align}
where in the last step we have used the composition rule~(\ref{appendix:composition_rule}). Inserting this expression in~(\ref{WignerAux}), we obtain
\begin{align}
W_{\ket{-1}}(\gamma)=\braket{ j,-j|\hat{D}(\bar{\gamma})\hat{\Delta}\hat{D}^\dagger(\bar{\gamma})|j,-j}.
\end{align}
Finally, using the expression~(\ref{appendix:kernel}) for the displaced kernel and taking into account Eqs.~(\ref{appendix:simplify_1}), together with
\begin{subequations}
\begin{align}
\braket{j,-j;l,m|j,-j}&=\delta_{m0}(-1)^l\braket{j,j;l,0|j,j},
\\
(-1)^lP_l(x)&=P_l(-x),
\end{align}
\end{subequations}
we obtain expression~(\ref{eq:coherent_terms_Wigner}) for $W_{\ket{-1}}(\gamma)$.

Let us now consider the Wigner function of the horizontal cat state $\ket{\psi_\text{H}}=\ket1+\ket{-1}$, which is expanded as
\begin{align}
W_{\ket{\psi_\text{H}}}(\gamma)=W_{\ket1}(\gamma)+W_{\ket{-1}}(\gamma)+2\text{Re}\big\{W_{\ket{-1}\bra{1}}(\gamma)\big\}.
\end{align}
Using various results from the previous lines, the last term is written as
\begin{align}
W_{\ket{-1}\bra{1}}(\gamma)&=\bra{1}\hat{D}(\gamma)\hat{\Delta}\hat{D}^\dagger(\gamma)\ket{-1}
\\
&=\braket{ j,j|\hat{D}(-1)\hat{D}(\gamma)\hat{\Delta}\hat{D}^\dagger(\gamma)\hat{D}(1)|j,-j}\nonumber
\\
&=\text{e}^{-2\text{i}j\bar\varphi}\braket{ j,j|\hat{D}(\bar{\gamma})\hat{\Delta}\hat{D}^\dagger(\bar{\gamma})|j,-j},\nonumber
\end{align}
with $\bar{\varphi}=-2\arg\{\gamma+1\}$ coming from the composition rule for the displacements. Inserting now the expression~(\ref{appendix:kernel}) for the displaced kernel, taking into account that
\begin{subequations}\label{appendix:simplify_2}
\begin{align}
\braket{j,-j;l,m|j,j}&=\delta_{l,2j}\delta_{m,2j}(-1)^{2j}\sqrt{\frac{2j+1}{4j+1}},
\\
Y_{ll}(\theta,\phi)&=\frac{(-1)^l}{2^l l!}\sqrt{\frac{(2l+1)!}{4\pi}}\text{e}^{\text{i}l\phi}\sin^l\theta,
\end{align}
\end{subequations}
and considering only the real part of $W_{\ket{-1}\bra{1}}(\gamma)$ we obtain precisely the $I_\text{H}(\gamma)$ term of Eq.~(\ref{eq:horizontalcat_SU(2)_interference}).

The same approach is easily applied to the vertical cat $\ket{\psi_\text{V}}=\ket{\text{i}}+\ket{-\text{i}}$. On the other hand, for the Wigner function of the compass state $\ket{\psi_\text{C}}=\ket{\psi_\text{H}}+\ket{\psi_\text{V}}$ we also need to evaluate cross terms between vertical and horizontal coherent states of the type 
\begin{equation}
\bra{\text{i}}\hat{D}(\gamma)\hat\Delta\hat{D}^\dagger(\gamma)\ket1.
\end{equation}
These cross terms can be obtained by using Eq.~(\ref{eq:SU(2)_superposition_Wigner}). However, numerical inspection of this expression shows that for $j\gg 1$ these terms do not contribute for small $|\gamma|$, as mentioned in the main text.

\bibliography{References}
\end{document}